\pdfoutput=1
\documentclass[structabstract,longauth]{aa}  
\usepackage{graphicx}
\usepackage{longtable}
\usepackage{multirow}
\usepackage{natbib}
\usepackage{txfonts}
\usepackage{fixltx2e}
\usepackage{wasysym}
\bibpunct{(}{)}{;}{a}{}{,}
\begin{document}
\renewcommand{\topfraction}{0.85}
\renewcommand{\bottomfraction}{0.7}
\renewcommand{\textfraction}{0.15}
\renewcommand{\floatpagefraction}{0.66}
\large
\def\g{\hbox{$\gamma$}}
\def\sig{\hbox{$\sigma$}}
\def\psrb{PSR~B1259$-$63}
\def\psrbss{PSR~B1259$-$63/SS2883}
\def\hessj{HESS~J1303$-$631}
\def\psrj{PSR~J1301$-$6305}
\def\flux{\textrm{\,TeV}^{-1}\textrm{\,cm}^{-2}\textrm{\,s}^{-1}}
\def\iras{IRAS~13010$-$6254}
\def\psrxmm{2XMM~J130145.7$-$630536}
\def\otherxmm{2XMM~J130234.9$-$630754}
\def\softxmm{2XMM~J130141.3$-$630535}
\def\testpsfxmm{2XMM\,J130138.3$-$630653}

\title{Identification of \hessj\ as a Pulsar Wind Nebula through $\gamma$-ray, X-ray and radio observations}
\offprints{Matthew L. Dalton, \email{dalton@cenbg.in2p3.fr}}
\large

\author{H.E.S.S. Collaboration
\and A.~Abramowski \inst{1}
\and F.~Acero \inst{2}
\and F.~Aharonian \inst{3,4,5}
\and A.G.~Akhperjanian \inst{6,5}
\and G.~Anton \inst{7}
\and S.~Balenderan \inst{8}
\and A.~Balzer \inst{7}
\and A.~Barnacka \inst{9,10}
\and Y.~Becherini \inst{11,12}
% ...
}

\institute{
Universit\"at Hamburg, Institut f\"ur Experimentalphysik, Luruper Chaussee 149, D 22761 Hamburg, Germany \and
Laboratoire Univers et Particules de Montpellier, Universit\'e Montpellier 2, CNRS/IN2P3,  CC 72, Place Eug\`ene Bataillon, F-34095 Montpellier Cedex 5, France \and
Max-Planck-Institut f\"ur Kernphysik, P.O. Box 103980, D 69029 Heidelberg, Germany \and
Dublin Institute for Advanced Studies, 31 Fitzwilliam Place, Dublin 2, Ireland \and
National Academy of Sciences of the Republic of Armenia, Yerevan  \and
Yerevan Physics Institute, 2 Alikhanian Brothers St., 375036 Yerevan, Armenia \and
Universit\"at Erlangen-N\"urnberg, Physikalisches Institut, Erwin-Rommel-Str. 1, D 91058 Erlangen, Germany \and
University of Durham, Department of Physics, South Road, Durham DH1 3LE, U.K. \and
Nicolaus Copernicus Astronomical Center, ul. Bartycka 18, 00-716 Warsaw, Poland \and
CEA Saclay, DSM/IRFU, F-91191 Gif-Sur-Yvette Cedex, France \and
APC, AstroParticule et Cosmologie, Universit\'{e} Paris Diderot, CNRS/ IN2P3,CEA/ lrfu, Observatoire de Paris, Sorbonne Paris Cit\'{e}, 10, rue Alice Domon et L\'{e}onie Duquet, 75205 Paris Cedex 13, France,  \and
Laboratoire Leprince-Ringuet, Ecole Polytechnique, CNRS/IN2P3, F-91128 Palaiseau, France
% ...
}

\abstract
{} 
%
{The previously unidentified very high-energy (VHE; $E > 100$\,GeV) $\gamma$-ray 
source \hessj, discovered in 2004, is re-examined including
new data from the H.E.S.S. Cherenkov telescope array in order to identify this object. 
Archival data from the \textit{XMM-Newton} X-ray satellite 
and from the PMN radio survey are also examined. 
}
{Detailed morphological and spectral studies of VHE $\gamma$-ray emission as well as of
the \textit{XMM-Newton} X-ray data are performed.
Radio data from the PMN survey are used as well to construct a leptonic model of the source.
The $\gamma$-ray and X-ray spectra and radio upper limit are used
to construct a one zone leptonic model of the spectral energy distribution (SED). 
}
%
{
Significant energy-dependent morphology of the $\gamma$-ray source is detected
with high-energy emission ($E > 10$\,TeV) positionally coincident with the
pulsar \psrj\ and lower energy emission ($E <2$\,TeV) extending $\sim 0.4^{\circ}$ to the
South-East of the pulsar. 
The spectrum of the VHE source can be described with a power-law with an exponential cut-off
$N_{0} = (5.6 \pm 0.5) \times 10^{-12}\flux$, 
$\Gamma = 1.5 \pm 0.2)$ and
$E_{\rm cut} = (7.7 \pm 2.2)$\,TeV.
The PWN is also detected in X-rays, extending $\sim 2-3'$ from the pulsar position 
towards the center of the $\gamma$-ray emission region. 
A potential radio counterpart from the PMN survey is also discussed, 
showing a hint for a counterpart at the edge of the X-ray PWN trail
and is taken as an upper limit in the SED. 
The extended X-ray PWN has an unabsorbed flux of 
$F_{\rm 2-10\,keV} \sim 1.6^{+0.2}_{-0.4}\times 10^{-13}\textrm{\,erg\,cm}^{-2}\textrm{\,s}^{-1}$ 
and is detected at a significance of $6.5\,\sigma$.
The SED is well described 
by a one zone leptonic scenario which, with its associated caveats, predicts a very low average
magnetic field for this source.
}
%
{
Significant energy-dependent morphology of this source, as well as the identification of an
associated X-ray PWN from \textit{XMM-Newton} observations enable identification of the VHE source
as an evolved PWN associated to the pulsar \psrj. This identification is supported by the one zone leptonic model, 
which suggests that the energetics of the $\gamma$-ray and X-ray radiation are such that
they may have a similar origin in the pulsar nebula. However, the large discrepancy in
emission region sizes and the low level of synchrotron radiation suggest a multi-population leptonic nature.
The low implied magnetic field suggests that the PWN has undergone significant expansion.
This would explain the low level of synchrotron radiation and the difficulty in 
detecting counterparts at lower energies, the reason
this source was originally classified as a ``dark'' VHE $\gamma$-ray source.
}
 
\keywords{Gamma-rays: observations -- Pulsars: individual: \psrj\ -- ISM: individual objects: \hessj}

\date{Received June 14, 2012; accepted October 12, 2012}

\authorrunning{A. Abramowski et al. (H.E.S.S. Collaboration)}
\titlerunning{Identification of \hessj\ as a pulsar wind nebula}

\maketitle

\section{Introduction}
In recent years, nearly a hundred VHE $\gamma$-ray sources have been discovered
by various experiments, including many different types of sources.
Generally, sources from these different classes
also exhibit radio and X-ray radiation, however, 
the discovery of TeV J2032+4130 by the HEGRA collaboration in 
2002 \citep{TeV2032} lead to a new class of extended 
Galactic VHE $\gamma$-ray sources without obvious counterparts at 
other wavelengths.  
\hessj, serendipitously discovered by H.E.S.S. 
(High Energy Stereoscopic System) during an observation campaign of the
pulsar \psrb\ \citep{psrb1259}, in observations taken between January and June of 2004 \citep{serendip}.
\hessj\ was the first so-called ``dark source'' discovered by H.E.S.S.
More of these sources were discovered by the H.E.S.S. 
collaboration in the following years \citep{unidentified,unidentified2}.  
Identifying and understanding this new class of sources has 
become an important task for modern $\gamma$-ray astronomy.

A growing number of extended VHE $\gamma$-ray sources, 
without (or with significantly fainter or less extended) X-ray/radio counterparts, 
appear to be associated with energetic pulsars in the Galactic Plane.  
Some recent examples of this class of objects include
HESS J1825$-$137 \citep{j1825}
and HESS J1356$-$645 \citep{j1356}.
These associations are 
believed to represent pulsar wind nebulae (PWNe), 
which are thought to be powered by a relativistic lepton-dominated
particle outflow from a central pulsar.
Many of these PWN associations form what are known as
\textit{Offset PWNe} where the pulsar is located at or near the edge
of the $\gamma$-ray and X-ray emission regions.  These configurations
may form in two ways. First, a high spatial velocity pulsar,
possibly supersonic (in which case a bow shock nebula may form), leaves behind
a ``trail'' of high-energy electrons in the ambient medium. 
Alternatively, an offset PWN may form if its expansion is blocked on one side
by the reverse shock of the supernova remnant (SNR) in which the pulsar was born.
Due to inhomogeneous densities in the ISM, the expansion of the supernova remnant
may proceed asymmetrically, or the motion of the pulsar may place it near the edge
of the SNR and the expanding PWN may then be disrupted asymmetrically by the reverse shock of
the SNR, a scenario known as a \textit{Crushed PWN} \citep{crushedpwn}.

At the time of discovery, \hessj\ was found to have a large intrinsic Gaussian 
extent of $\sim 0.16^{\circ}$, assuming a 2-dimensional symmetric
Gaussian distribution, and a flux of $\sim 17$\% of the Crab flux above 380 GeV.
Originally, the source had no known extended counterparts at other wavelengths
and was, therefore, classified as a dark source. As is the case with many such dark sources,
\hessj\ is found to have a pulsar lying near the edge of the emission region with a high enough
spin-down luminosity to account for the $\gamma$-ray emission. \psrj\ is located at the
northwestern edge of the emission region of the H.E.S.S. source and,
with a spin-down luminosity $\dot{E} = 1.7\times 10^{36}$\,erg\,s$^{-1}$, 
is the most powerful pulsar within $6^{\circ}$ of the H.E.S.S. source 
(\cite{atnfcat}, see Table~\ref{table:pulsars} for a list of known pulsars within $0.5^{\circ}$ of
\hessj). This pulsar is young, with a characteristic age of $\tau_{c} = 11$
kyr, and a rotation period of 184\,ms.  

Originally, the distance to \psrj\ was estimated to be 15.8\,kpc, based on \cite{oldgalne},
which would have required a rather high $\gamma$-ray conversion efficiency of 37\% in the 0.3 to $10\,$TeV range.
Using a newer model of the Galactic electron distribution, NE2001 \citep{galne}, however, yields a much closer
distance of $6.6\,$kpc. 
Using this updated distance estimate, the reported VHE $\gamma$-ray spectrum at the time of
discovery yields an integrated flux in the 1 to 30\,TeV band of $\Phi = 1.2\times 10^{-11}$\,erg\,cm$^{-2}$\,s$^{-1}$
or 3.7\% of the current spin-down luminosity of this pulsar, 
($F_{6.6} = \dot{E}/4\pi(6.6\textrm{ kpc})^{2} = 3.26\times 10^{-10}$\,erg\,cm$^{-2}$\,s$^{-1}$), 
a $\gamma$-ray conversion efficiency which is comparable to other VHE PWNe (typically 0-7\%, see e.g. \cite{gammax}).
\begin{table}[p]
\caption{All known pulsars within $0.5^{\circ}$ of \hessj\ \citep{atnfcat}.
$\delta_{10\mathrm{\,TeV}}$ is the distance from the given pulsar to the $E > 10$\,TeV peak position.}
\begin{tabular}{ l l r r}
Pulsar & $\dot{E}/10^{30}$\,erg/s & $\delta_{10\mathrm{\,TeV}}$ [arc min]\\
\hline
PSR J1301$-$6305 & 1,700,000 & 3.1 \\
PSR J1301$-$6310 &     6,800 & 6.6 \\
PSR J1305$-$6256 &       760 & 24.2 \\
PSR J1302$-$6313 &       270 & 7.2 \\
PSR J1303$-$6305 &         7 & 5.6 
\end{tabular}
\label{table:pulsars}
\end{table}

A  5 ksec \textit{Chandra} X-ray observation, partially covering the VHE peak emission
region \citep{chandra}, revealed several point sources within the field of view, but no extended 
emission corresponding to the $\gamma$-ray emission region 
was found, and none of the radio pulsars in the field of view of the \textit{Chandra}
observation were detected. The possibility of an annihilating clump of 
dark matter as the origin of the $\gamma$-ray signal was explored by \cite{darkmatter}.
Such a model could explain the lack of detection of lower energy counterparts.
However, it was found that the spectrum obtained for this source would 
require an unreasonably high mass for the candidate dark 
matter particles ($\sim 40$\,TeV).  Also, as mentioned in that study, the inferred lateral 
density distribution does not support a dark matter scenario.
\cite{darkmatter}, therefore, concluded this to be an
unlikely candidate for the explanation of the VHE source.

To build a complete picture of the $\gamma$-ray emission process in this source, 
data from recent re-observations of \hessj\ with the H.E.S.S. telescope array were analysed,
enabling studies of energy-dependent morphology. Also, follow-up observations by the
\textit{XMM-Newton} X-ray satellite, performed in 2005, showing a detection of a compact 
source slightly offset from the pulsar position and a significantly extended PWN,
are presented.

In Section 2, the H.E.S.S. instrument, data and analysis methods are discussed
as well as the light curve.  
Section 3 describes the studies of energy-dependent morphology of the
$\gamma$-ray source, followed by a discussion of the spectrum 
of the source in Section 4. Section 5 presents the results of the
\textit{XMM-Newton} X-ray follow-up observations, showing an X-ray PWN
associated with the pulsar \psrj.
Finally, in Section 6, the implications of the analysis are discussed and the 
case is made for an association of \hessj\ with the pulsar \psrj.

\section{H.E.S.S. observations and analysis}
\subsection{The H.E.S.S. instrument}
H.E.S.S. is an array of four imaging 
atmospheric Cherenkov telescopes
located in the Khomas Highland of Namibia ($23^{\circ}16\arcmin18\arcsec$\,S,
$16^{\circ} 30\arcmin 00\arcsec$\,E) at an altitude of 1800\,m above sea-level. 
The telescopes image the Cherenkov light emitted by charged
particles in the extensive air shower created when a $\gamma$-ray is absorbed
in the atmosphere. 
They are optimized for detection of VHE $\gamma$-ray initiated showers in the energy range of
hundreds of GeV to tens of TeV by 
Each telescope has a $107\,\textrm{m}^{2}$ tessellated mirror
surface and is equipped with a 960 photomultiplier tube camera with a field of
view (FoV) diameter of $\sim 5^{\circ}$ \citep{HESS,HESS2}. The telescopes are
triggered in coincidence mode \citep{Funk} assuring that an event is always
recorded by at least two of the four telescopes allowing stereoscopic
reconstruction of the showers. More information about H.E.S.S. 
can be found in \cite{hinton}. 

\subsection{Data and analysis techniques}
\label{data}
\hessj\ was originally discovered during an observation campaign for \psrb.  
Follow-up observations of the two sources between 2004 and 2008 led 
to a total dataset of 108.3 hours of live time, using only observations which passed 
standard H.E.S.S. data quality selection which rejects observations taken during periods of bad weather
or with instrumental irregularities.
The data were taken in wobble mode at an average zenith angle of $43.8^{\circ}$,
with an average offset of $0.8^{\circ}$ from the
position reported in the discovery paper \citep{serendip}.

The data were analyzed using H.E.S.S. standard Hillas reconstruction \citep{crab}.
Cuts were applied to the shower image parameters to minimize background, primarily due to cosmic-ray
protons. For spectrum extraction, standard cuts (also defined in \cite{crab}), were used
together with the Reflected-Region Background method \citep{bg} to subtract residual cosmic ray
background, which resulted in an average energy threshold of $\sim 720$ GeV.  The resulting
excess for this analysis was found to be  12085 photons for a detection significance of $33\,\sigma$.
Some of these early observations were made
with telescope pointings coincident with the \hessj\ emission region, rendering them unsuitable 
for spectral analysis since placement of reflected regions for background estimation is not possible.  
For the morphology studies, hard cuts were applied to further reduce background contamination
and improve image reconstruction, and hence the point spread function (PSF) of
the instrument, at the expense of a higher energy threshold, together with the Ring Background method,
resulting in an average energy threshold of $\sim 840$ GeV.
Cross-checks were performed using a multi-variate analysis \citep{ohm}, with background 
suppression based on boosted-decision trees, leading to compatible results.

\subsection{VHE $\gamma$-ray map and light curve}
\begin{figure}
\centering
\resizebox{\hsize}{!}{\includegraphics{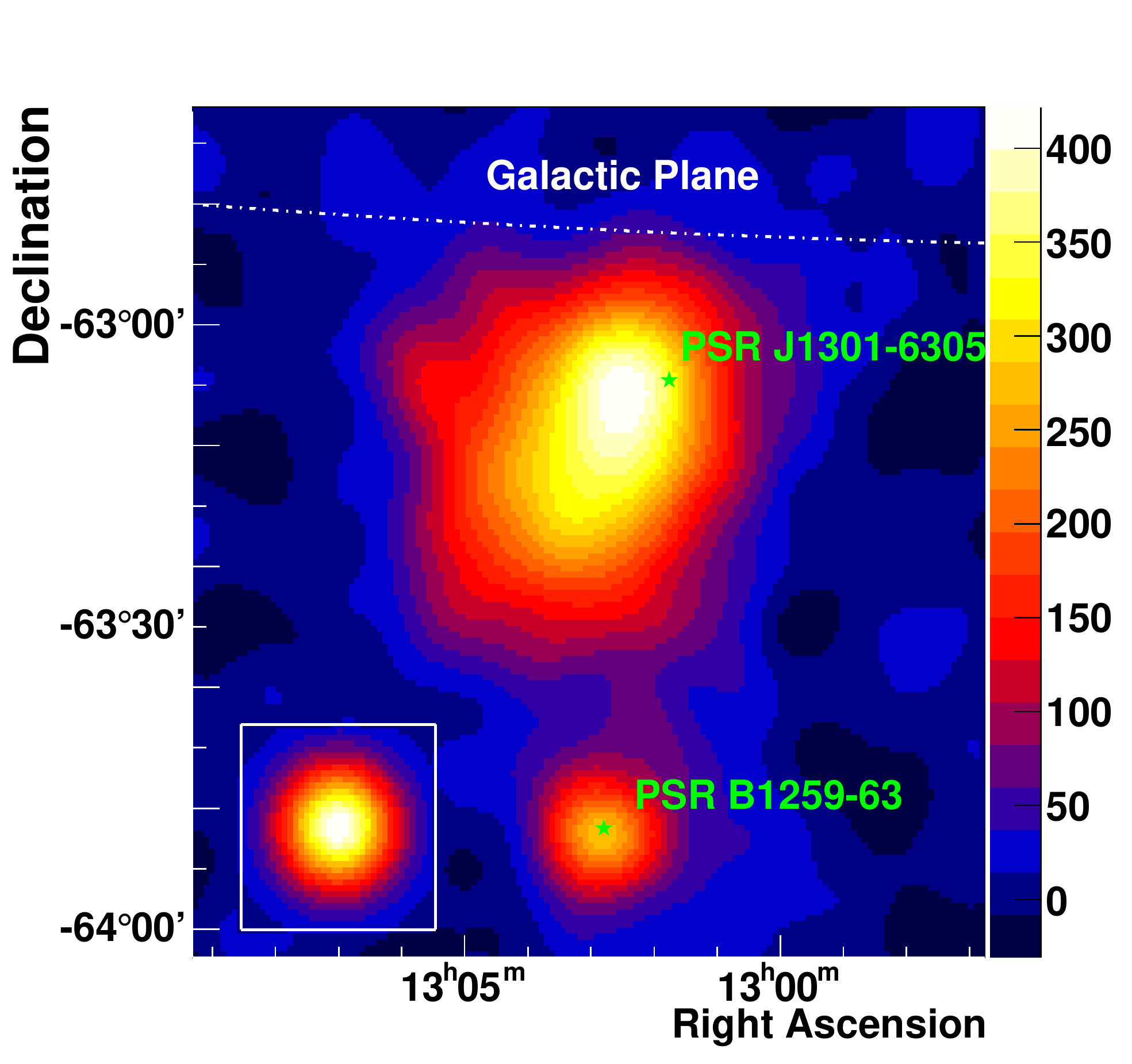}}
\caption{The \hessj\ VHE $\gamma$-ray excess map, produced using hard cuts and the ring
background method, was smoothed with a Gaussian kernel with
$\sigma = 0.05^{\circ}$.  Coordinates are J2000.0.
The high spin-down power pulsar, \psrj, is indicated with a green star 
to the North-West of \hessj.
The point source, associated to the pulsar 
\psrb, is seen in the bottom of the FoV. The size of the
H.E.S.S. PSF, also smoothed with a Gaussian kernel with
$\sigma = 0.05^{\circ}$, is shown in the white box to the lower left.
The blue/red transition occurs at a detection significance of $\sim 5\sigma$.
}
\label{skymap}
\end{figure}

The VHE $\gamma$-ray excess map (Fig.~\ref{skymap}) of the \hessj\ FoV shows
extended emission to the South-East of \psrj. A fit of 
a two-dimensional asymmetric Gaussian function 
to the excess resulted in a best-fit position of $\alpha
= 13^{\textrm{h}}02^{\textrm{m}}48^{\textrm{s}} \pm
3^{\textrm{s}}_{\rm{stat}}$, $\delta = -63^{\circ}10'39'' \pm
24''_{\rm{stat}}$ (J2000.0), with major/minor axis Gaussian widths
of $\sigma_{\rm x} = 0.194^{\circ} \pm 0.008^{\circ}$ and $\sigma_{\rm y} =
0.145^{\circ} \pm 0.006^{\circ}$, with a position angle (counter clockwise from north) of $\phi = 147^{\circ}
\pm 6^{\circ}$. The $\chi^{2}/NDF$ of the fit was 390 / 391.
The exposure gradient over the source extension was found to be small and have a negligible affect
on the resulting source position.  The fitted position is consistent with the one quoted in the original discovery paper \citep{serendip}, but 
slightly shifted towards the pulsar position due to the (compared to the 
discovery paper) higher energy threshold of the hard cuts used and the 
presence of energy-dependent morphology (see Section 3). 

The nightly flux was determined using a flux extraction region of radius $0.6^{\circ}$ to ensure full
enclosure of the source, around the best fit position given above assuming a power-law spectrum
with an index of 1.5.  Studies were performed to account for influences from the nearby VHE
source \psrb. As expected for an extended source, with an estimated diameter of 40\,pc at a distance of 6.6\,kpc, 
the nightly flux is consistent with constant emission, with $\chi^2/NDF = 77/69$,
verifying the stability of the H.E.S.S. instrument over the period of data taking.


\section{Energy-dependent morphology}

\begin{figure}
\centering
\resizebox{\hsize}{!}{\includegraphics{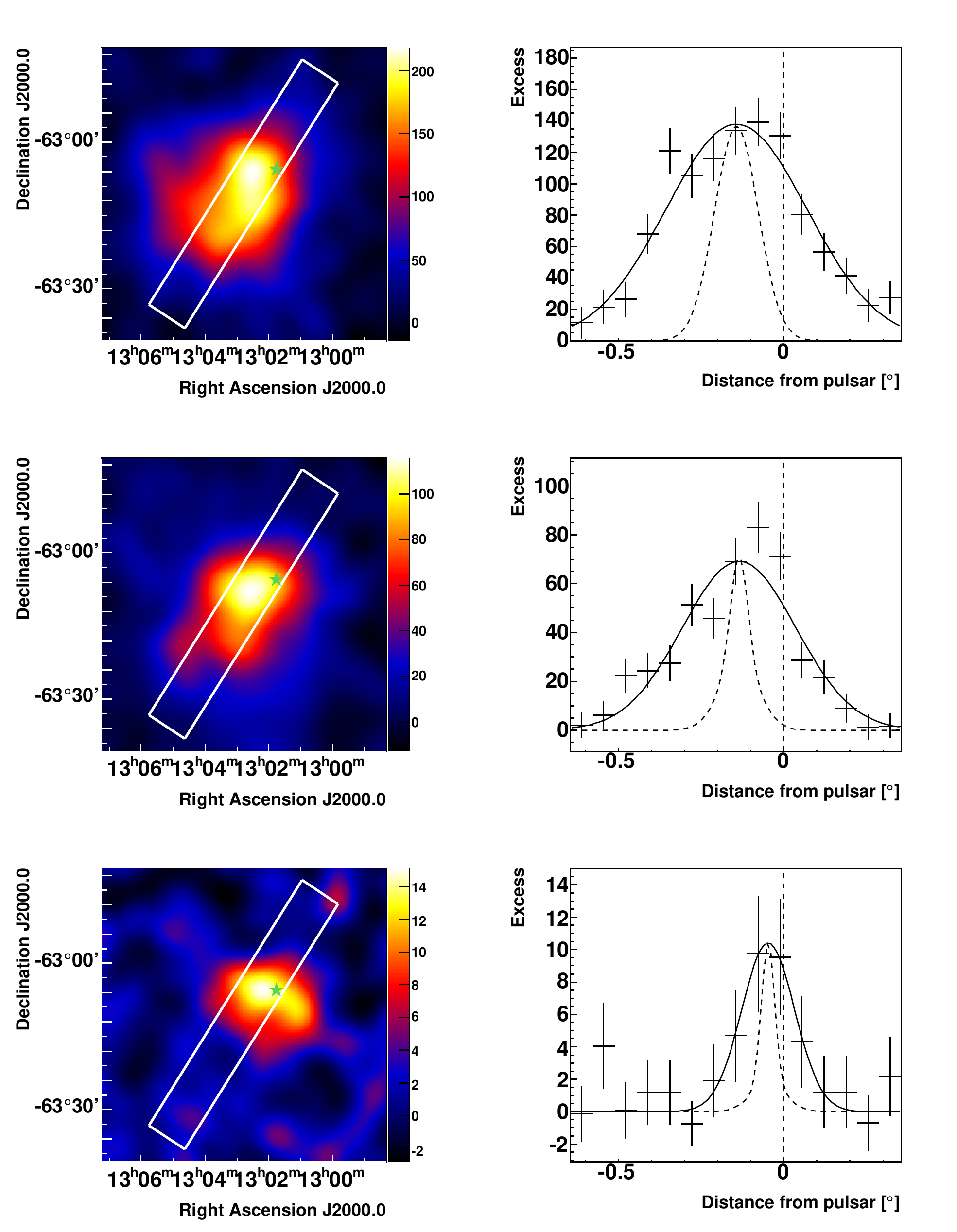}}
\caption{Left: uncorrolated excess images of the \hessj\ region in the energy
bands 
E$_{1}$ = (0.84 - 2)\,TeV, E$_{2}$ = (2 - 10)\,TeV and
E$_{3} >10$\,TeV (from top to bottom). Coordinates are J2000.0. All images were smoothed
with a Gaussian kernel of width $0.05^{\circ}$. 
Slices are indicated by the rectangles,
taken in the direction of the semi-major axis of the fitted asymmetric Gaussian function.
Right: the slices on the uncorrelated excess images are then fitted with a Gaussian function.  The
pulsar position is marked by a green star in the sky maps and a dashed line in the profiles.
The dashed curves show the energy-dependent PSF of the H.E.S.S. instrument.}
\label{edepskymaps}
\end{figure}

To test for the presence of energy-dependent morphology in the VHE source, 
excess images were generated in the following energy bands:
E$_{1}$ = (0.84 - 2)\,TeV, E$_{2}$ = (2 - 10)\,TeV and
E$_{3} >10$\,TeV (Fig. \ref{edepskymaps}, left, top to bottom).
The radial acceptance of the FoV was determined from the data, thus naturally accounting for the
energy dependence.
Slices were made on the uncorrelated excess images having dimensions of 
$1.0^\circ \times 0.1^\circ$ and centered at the
best fit position of the VHE excess. The orientation is chosen along the
fitted position angle (see Sec.~\ref{data}). A Gaussian function was then fit to
each slice as shown in Fig.~\ref{edepskymaps} (right).  
The intrinsic source width was obtained by fitting the convolution of a Gaussian with
the energy dependent H.E.S.S. PSF.
 
\begin{table}[p]
\caption{Results of the Gaussian fit to the  slices on the excess images in the energy bands 
E$_{1}$ = (0.84 - 2)\,TeV, E$_{2}$ =(2 - 10)\,TeV and
E$_{3} >10$\,TeV.
$c$ is the mean of the Gaussian, $w_{\rm img}$ is the Gaussian width and
$w_{\rm int}$ is the intrinsic Gaussian width of the source after correcting for the
PSF, P is the p-value of the $\chi^{2}$ fit.}
\begin{tabular}{ r l l l l}
Band& $c$ & $w_{\rm img}$ & $w_{\rm int}$ & P \\
\hline
$E_{1}$ & $ -0.14^{\circ} \pm 0.01^{\circ} $  &  $ 0.21^{\circ} \pm 0.01^{\circ} $  &  $
0.20^{\circ} \pm 0.01^{\circ} $  & 0.42 \\ 
$E_{2}$ & $ -0.13^{\circ} \pm 0.01^{\circ} $  &  $ 0.17^{\circ} \pm 0.01^{\circ} $  &  $
0.16^{\circ} \pm 0.01^{\circ} $  & 0.14 \\ 
$E_{3}$ & $ -0.05^{\circ} \pm 0.02^{\circ} $  &  $ 0.08^{\circ} \pm 0.02^{\circ} $  &  $
0.07^{\circ} \pm 0.02^{\circ} $  & 0.97 \\
\end{tabular}
\label{table:morphfit}
\end{table}

\begin{table}[p]
\caption{Quality of fit of a constant vs. a line to the 
source intrinsic Gaussian extension and mean, measured from the pulsar
position, as a function of energy. The much improved p-values of the linear fits as compared to the
constant fits indicate the presence of significant energy-dependent morphology.}
\begin{tabular}{ r | l l l l}
Constant Fit& $\chi^{2}/NDF$ & p-value \\
\hline
$w_{\rm int}$ & 44.9 / 2  & $1.8\times 10^{-10}$ \\
$c$ & 18.3 / 2  & $1.1\times 10^{-4}$ \\
\hline
Linear fit\\
\hline
$w_{\rm int}$ & 0.4 / 1  & $0.55$ \\
$c$ & 2.4 / 1  & $0.12$
\end{tabular}
\label{table:morphfitq}
\end{table}

\begin{figure}
\centering
\resizebox{\hsize}{!}{\includegraphics{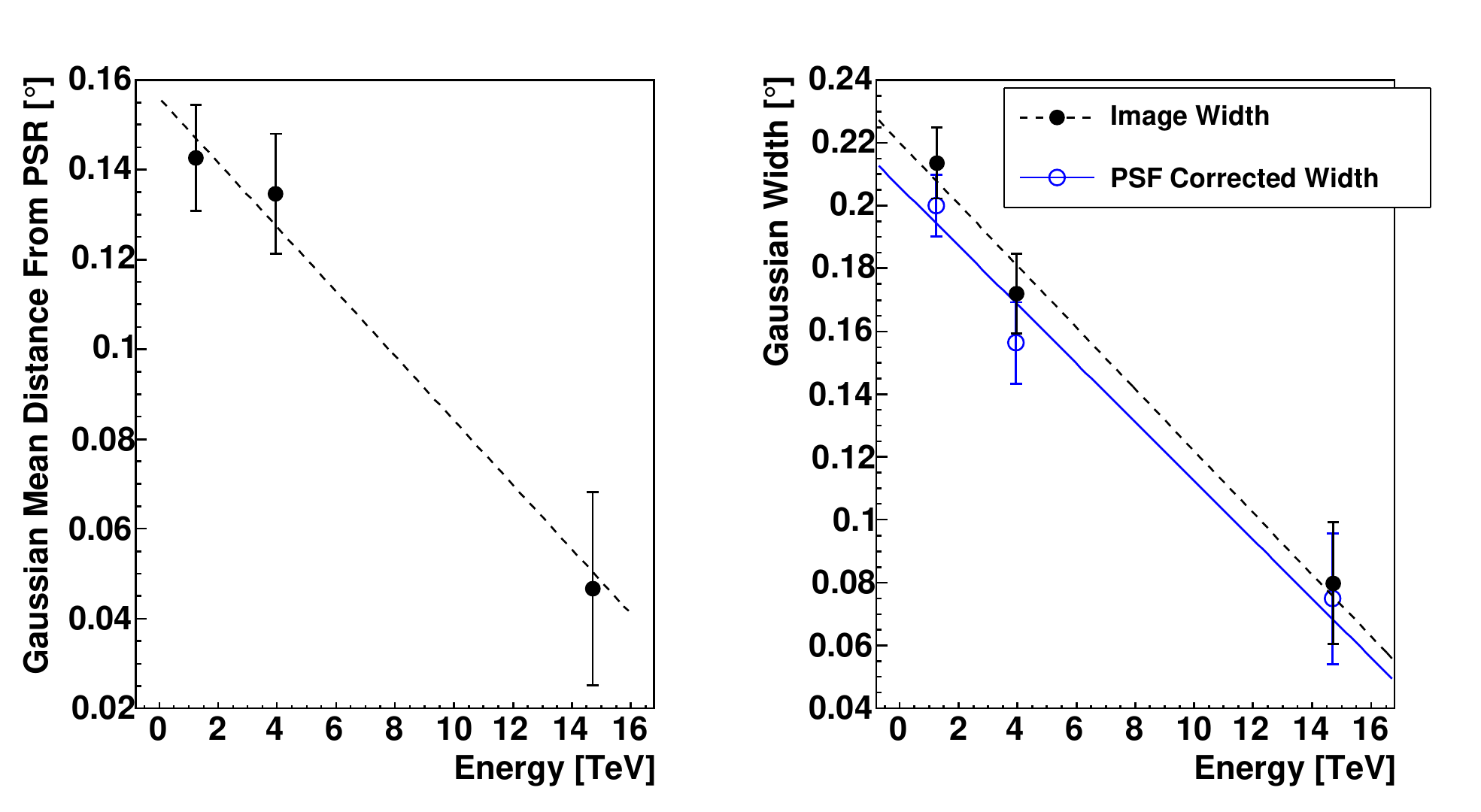}}
\caption{Left: \hessj\ fitted Gaussian mean, $c(E)$, measured from the pulsar
position, as a function of energy. Right: the PSF corrected intrinsic Gaussian extension
($w_{\rm int}(E)$, blue) is overlaid with the fitted uncorrected excess Gaussian extension
($w_{\rm img}(E)$, black dashed) as a function of energy.
The points are placed at the average energy of the photons falling in the corresponding
energy bin (indicated by the horizontal error bars, not used in fit).}
\label{gausfits}
\end{figure}

The resulting parameters of the PSF convolved Gaussian fits, mean $c$ and intrinsic
Gaussian width $w_{\rm int}$, for each energy band (Table~\ref{table:morphfit}) were then plotted
as a function of energy (Fig. ~\ref{gausfits}). A fit of a constant to these parameters yielded
very bad quality fits, which establishes the existence of strong energy-dependent morphology.
This morphology implies a spectral steepening in $\gamma$-rays away from the pulsar, a physical property
predicted to be present in evolved PWNe.
Fitting a linear function yielded much better quality fits (Tab.~\ref{table:morphfitq}) and a model
of the morphology parameterized by a projected center of emission, $c(E)$, calculated with respect
to the pulsar position,
and an intrinsic source Gaussian width, $w_{\rm int}(E)$, which is calculated by taking into account
the (energy-dependent) finite angular resolution of the instrument:
\begin{eqnarray*}
c &=& (0.157 \pm 0.012)^{\circ} - (0.006 \pm 0.002)^{\circ}\times E_{\rm\,TeV}\\ 
w_{\rm int} &=& (0.215 \pm 0.012)^{\circ} - (0.009 \pm 0.002)^{\circ}\times E_{\rm\,TeV} 
\end{eqnarray*}

\section{Energy spectrum}
\begin{figure}
\centering
\resizebox{\hsize}{!}{\includegraphics{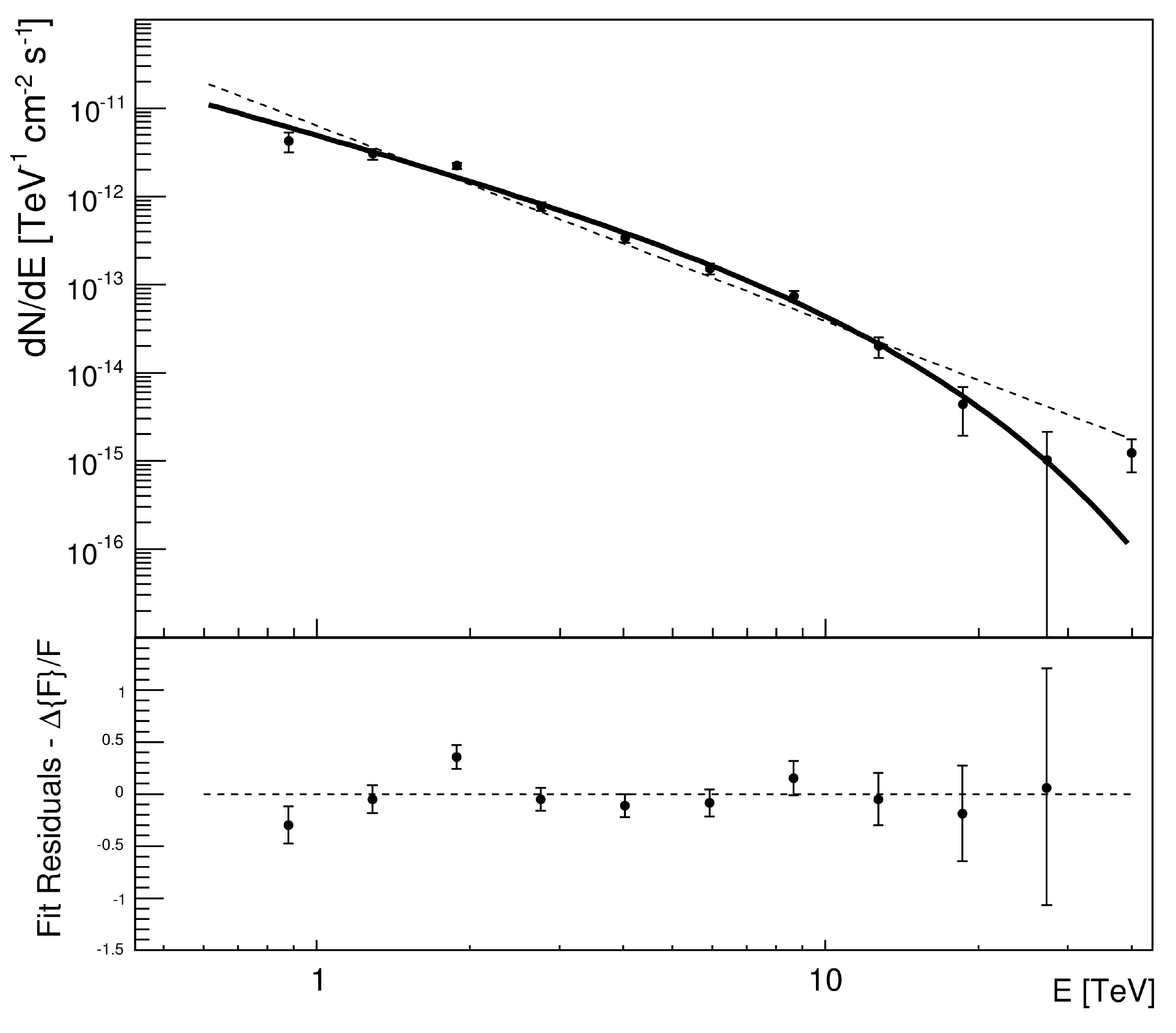}}
\caption{\hessj\ spectrum derived using an integration region of radius $0.6^{\circ}$.
The spectrum is well fit with a power-law function with spectral index 
$\Gamma = 1.5 \pm 0.2$ with a fitted cutoff energy of $E_{\rm cut} = (7.7 \pm 2.2)$\,TeV.
The fit resulted in a $\chi^{2}/NDF$ of (20 / 8) corresponding to 1\% p-value.
A fit of a power-law spectrum with no cutoff, shown by the dashed line, resulted in a p-value of
$7\times 10^{-8}$.
The last spectrum point deviates $\sim 2 \sigma$ from the fitted curve. It was removed from the
residuals plot for better visibility.}
\label{spectrum}
\end{figure}

The spectrum was derived using the Reflected-Region background method
with an integration region of radius $0.6^{\circ}$, roughly three times 
the intrinsic Gaussian extent at low energies to avoid effects of energy-dependent
morphology, centered at the fitted source position.
The derived spectrum for the entire dataset, excluding observations 
where the offset of the pointing position to the center of the source is less than $0.6^{\circ}$
(reducing the total live time to 70.3 hours), is shown in Fig.~\ref{spectrum}. The spectrum was fit
with a  power-law function, $dN/dE = N_{0}(E/1\mathrm{\,TeV})^{-\Gamma}$, with a resulting photon
index of $\Gamma = 2.44 \pm 0.03_{\rm{stat}}$ and a normalization constant $N_{0} = (5.9 \pm 0.3_{\rm{stat}})
\times 10^{-12}$ $\flux$. This normalization is larger than that found in the original discovery
paper due to a larger integration region. 
However, with the inclusion of the additional data taken since the source discovery, the
p-value of a chi-squared minimization is rather poor ($7\times 10^{-8}$). A chi-square fit to a power-law spectrum with 
a cut-off at the energy $E_{\rm cut}$, 
$$
\frac{dN}{dE} = N_{0}E^{-\Gamma}e^{-E/E_{\rm cut}},
$$
yielded a better p-value of 1\%, with fitted parameters 
$N_{0} = (5.6 \pm 0.5_{\rm stat}) \times 10^{-12}\flux$, 
$\Gamma = 1.5 \pm 0.2_{\rm stat}$ and
$E_{\rm cut} = (7.7 \pm 2.2_{\rm stat}$)\,TeV.
This spectrum yields an integrated flux in the $1-30$\,TeV band of $(2.5 \pm 0.1)\times
10^{-11}$\,erg\,cm$^{-2}$\,s$^{-1}$ or 7.7\% of $F_{6.6}$.
Monte-Carlo studies were preformed to test for a  possible
contribution from the source \psrb\ (spill over events), due to the position and 
size of the integration region and the exclusion region for \psrb.
Effects from this source are estimated to be about 2\% on the integrated flux,
smaller than statistical and systematic errors.

\section{\textit{XMM-Newton} X-ray observations}
In a search for counterparts of the VHE $\gamma$-ray source in the keV energy band,
two \textit{XMM-Newton} observations, each about 30 ksec, were carried out on 
July $12^{\rm th}$ and $14^{\rm th}$, 2005, in satellite 
revolution number 1024 (ObsID 0303440101, ``Observation 1'') and 
revolution 1025 (ObsID 0302340101, ``Observation 2'') respectively.
All three X-ray imaging CCD cameras (EPIC MOS1, MOS2, and pn) were operated in full-frame mode, 
with a medium filter to screen out optical and UV light,
with the exception of the pn camera during the first observation, where the 
Large Window mode with the Thin1 filter was used.

\subsection{Data analysis}
\begin{figure}[h!]
\centering
\resizebox{\hsize}{!}{\includegraphics{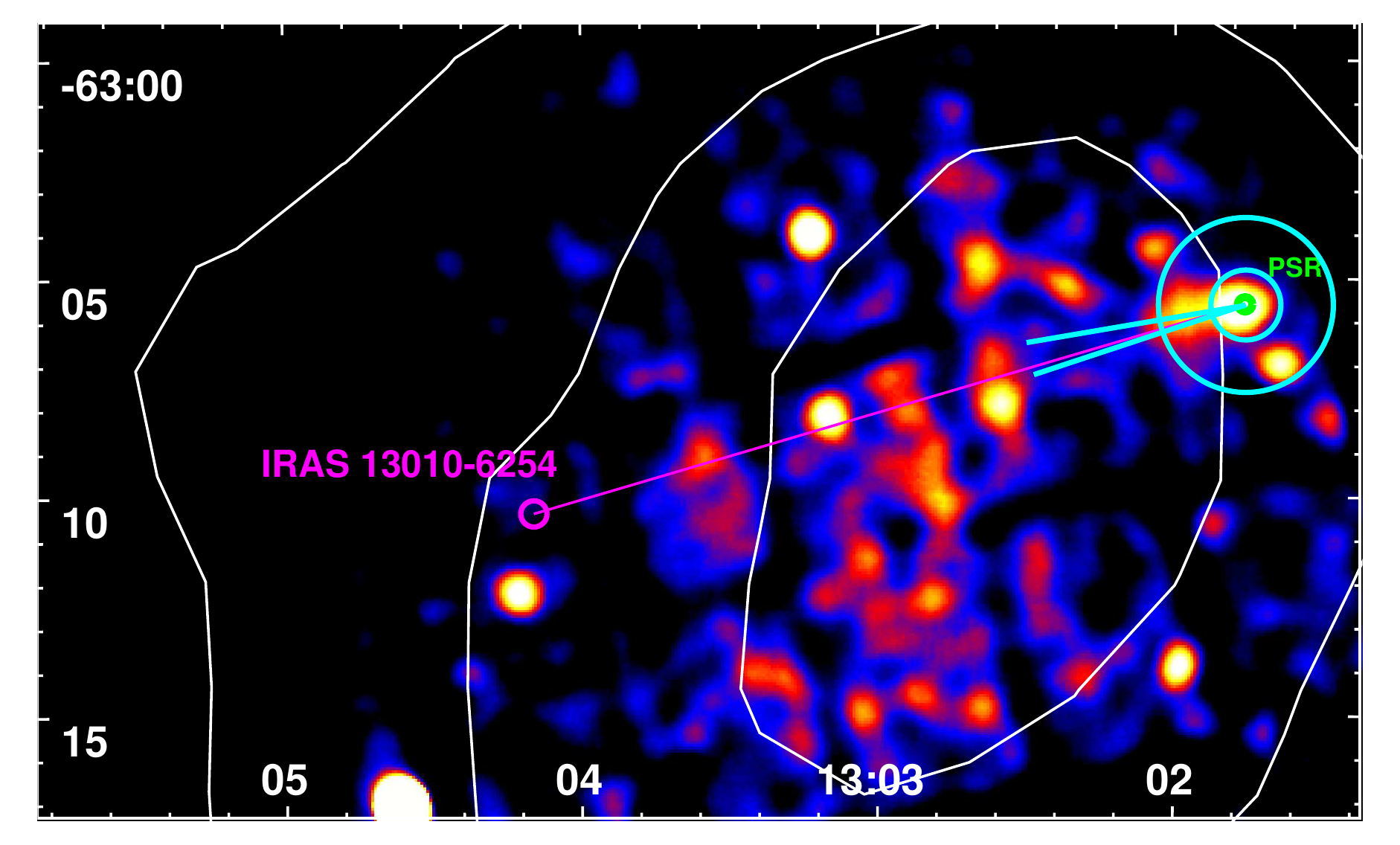}}
\resizebox{\hsize}{!}{\includegraphics{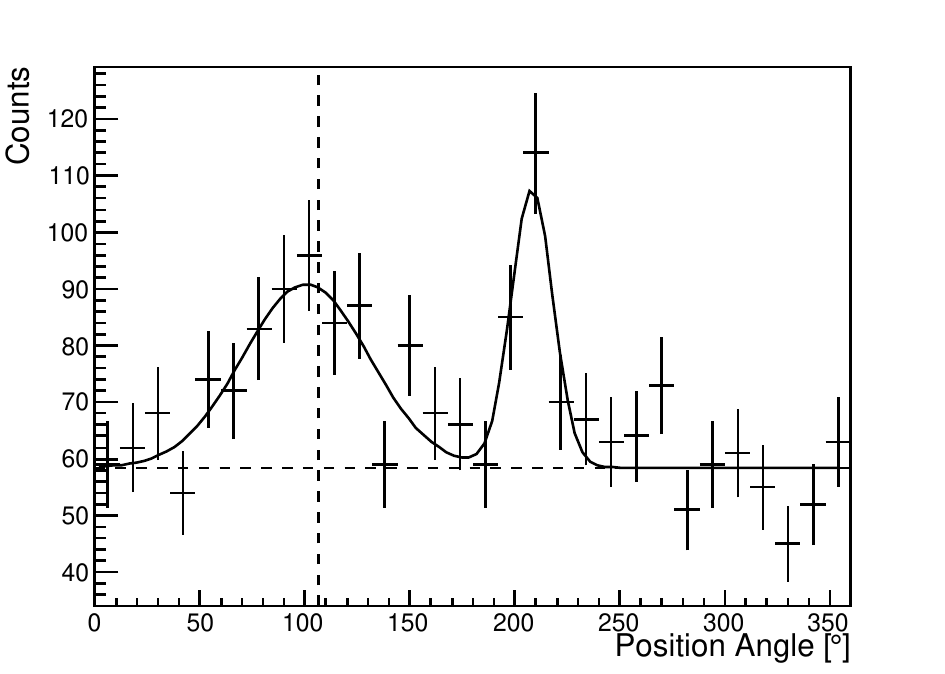}}
\caption{Top: The 2 - 8\,keV \textit{XMM-Newton} X-ray count map in the region of the pulsar smoothed by a
Gaussian kernel of width $\sigma=40''$.  The horizontal axis is Right Ascension and 
vertical axis is Declination in J2000.0 coordinates. 
The count map is not exposure corrected, thus the apparent enhanced emission near the center
is only an artifact.
Overlaid is the projection annulus, shown in cyan, used to determine the
direction of the X-ray extension, with inner radius of $48''$ and outer radius of $120''$, centered on the pulsar position.  
The 8, 14 and 20 $\sigma$ TeV significance contours are shown in white.
\newline
The direction of the 
extension is found to be within 1\,$\sigma$ from the direction of the star formation region, \iras, 
located at a position angle of $106.3^{\circ}$, indicated by a magenta line in the sky map and a
dashed line in the projection. The cyan lines in the sky map show the 1\,$\sigma$ errors in the fitted
direction of the extension.
Bottom: The X-ray azimuthal projection in position angle from the pulsar location.
The projected on-counts were fitted with the sum of a 
Gaussian and a flat background giving a position angle of $101.5^{\circ} \pm 5.3^{\circ}$
and a Gaussian width of $30^{\circ} \pm 7^{\circ}$.  The point source located at $\sim 210^{\circ}$
is an unidentified X-ray source.
}
\label{annulus}
\end{figure}
For the data analysis of these observations, the \textit{XMM-Newton} 
Science Analysis Software (SAS), version 9.0, was used (http://xmm.esac.esa.int/sas/).
Cleaning the data and removing periods of high background due to soft proton flares resulted in a combined data 
set of about 52 ksec exposure.  
For this analysis, the energy band 2 - 8\,keV was used to optimize the signal-to-noise ratio, since few events are
expected at lower energies due to high absorption.
The SAS task emosaicproc was used to combine the observations and perform source
detection, resulting in the detection of 73 point sources within the combined field of view above
the maximum likelihood threshold of 10.  The X-ray PWN associated to \psrj\ was also detected in this way, 
corresponding to the 2XMMi catalog source \psrxmm\ \citep{xmmcat},
with a flux of $F_{\rm 2-12\,keV} = (7.7 \pm 1.0) \times 10^{-14}\textrm{\,erg\,cm}^{-2}\textrm{\,s}^{-1}$
but peaked $15'' \pm 1.6''$ to the East of the pulsar with an extension of $6''$ at a maximum likelihood of 7.7
(sources with likelihood $<8$ may be spurious).
No emission corresponding to the full extension of the H.E.S.S. source was found.
However, a small apparently extended asymmetric X-ray source,
directly adjacent to \psrj, is seen extending roughly 
$2'$ from the pulsar position
towards the center of the VHE
$\gamma$-ray emission region (Fig.~\ref{annulus}). A detailed analysis of this feature is 
presented in the following section.

While Observation 1 has the pulsar position closer to on-axis than Observation 2, 
it is unfortunately not suited for studying the extended X-ray source since
the extended region found in Observation 2 lies directly 
on/between the edges of the CCD chips in all three detectors in Observation 1, 
thereby obscuring the view of this feature.
Therefore, only Observation 2 was used for further analysis.

\subsection{Extended X-ray PWN}
To determine the direction of the X-ray feature, possibly associated to \psrj,
an annular projection was taken around the pulsar position with
an inner radius of $48''$ and an outer radius of $120''$ (Fig.~\ref{annulus} top).
The projected counts were fitted with the sum of a 
Gaussian and a flat background giving a position angle of $101.5^{\circ} \pm 5.3^{\circ}$
and a Gaussian width of $30^{\circ} \pm 7^{\circ}$ (Fig.~\ref{annulus} bottom).  
The statistics were too low to warrant individual examination of the three cameras.
The direction of the extension as determined here was used for the orientation of the slice on the
count map, as presented below.

The direction of the X-ray extension is consistent to within 1\,$\sigma$ with the direction of
the star formation region \iras, on the opposite side but within 
the 8\,$\sigma$ significance contour of the VHE source, as indicated in Fig.~\ref{annulus} top.
This potential birthplace for the pulsar is considered in more detail in Sec.~\ref{discussion}.

\begin{figure}
\centering
\resizebox{\hsize}{!}{\includegraphics{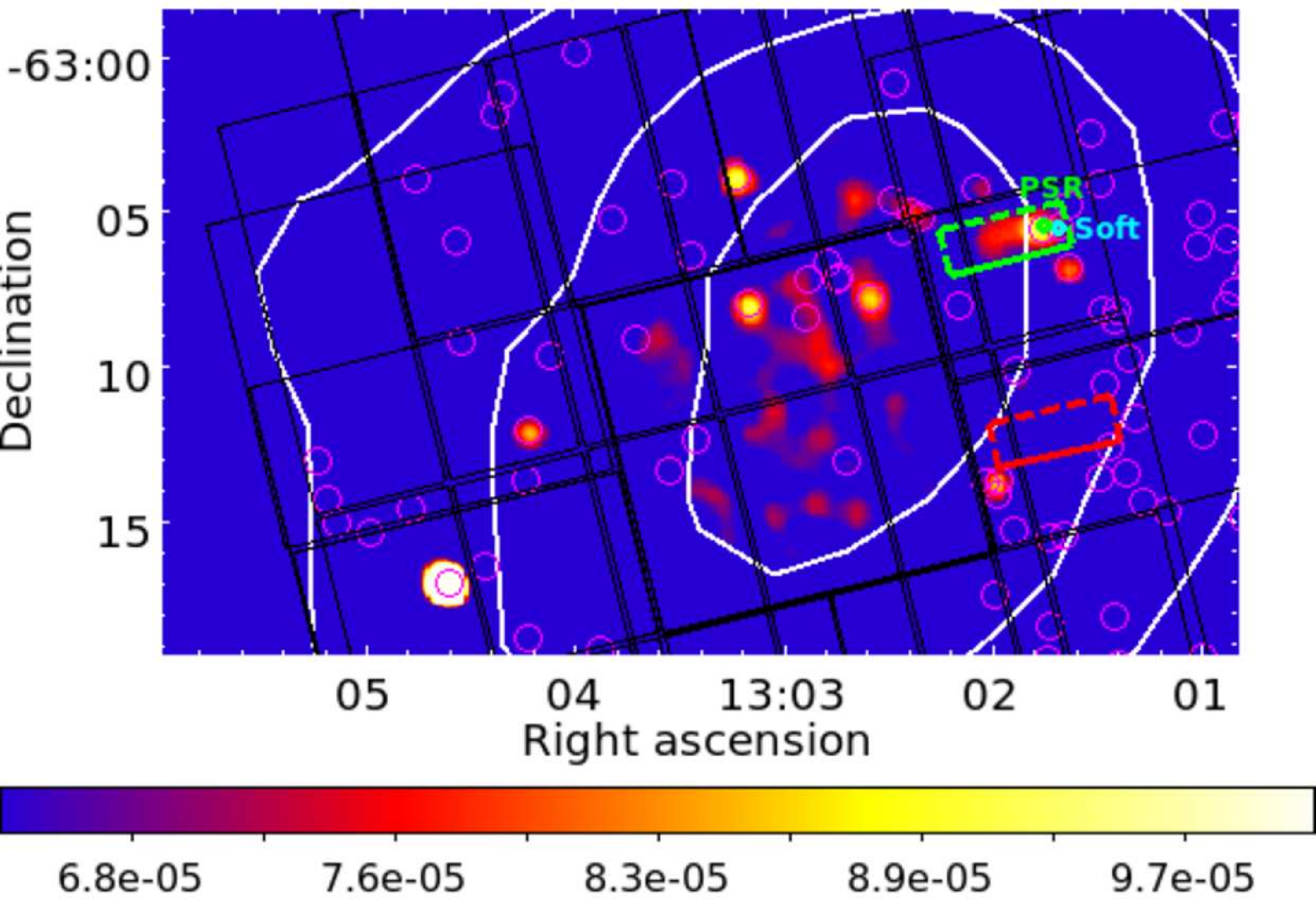}}
\resizebox{\hsize}{!}{\includegraphics{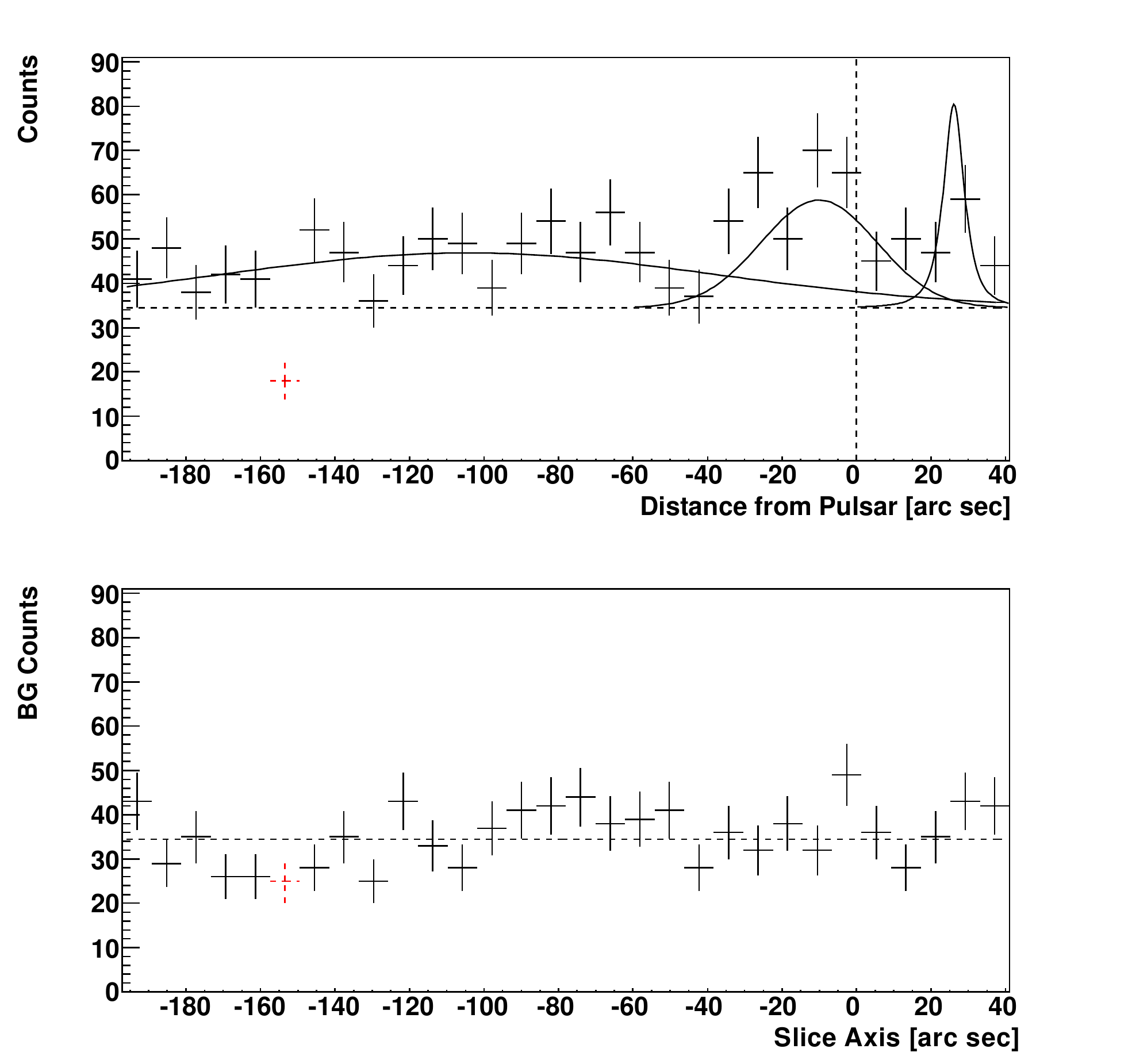}}
\caption{
Smoothed, exposure corrected XMM-Newton X-ray flux map (cm$^{-2}$\,s$^{-1}$) in the $2-8$\,keV 
energy band (top)
The horizontal axis is Right Ascension and the vertical axis is Declination
in J2000.0 coordinates. 
An extended X-ray source
appears from the position of the pulsar, \psrj\ (green dot), and extends roughly in the direction
of the center of the H.E.S.S. source (white contours show 8, 14 and 20\,$\sigma$ $\gamma$-ray significance). 
Chip edges are shown in black
for all three detectors and XMM catalog sources are shown in magenta. The green box shows the slice used
to create the profile (middle) and the red box shows the slice used for background estimation
(bottom). Both slices were taken from the un-exposure corrected count map.  
A presumably unrelated soft point source, \softxmm, is shown in the map as a cyan dot. 
The excess was simultaneously fit with two Gaussians (the ``compact region'' near the pulsar
position plus the ``diffuse region'' left of the pulsar) a King profile for the unrelated soft
source to the right of the pulsar position and a constant background from the bottom slice.
The red dotted bins in the projections
lie directly on a chip edge in the pn camera in both slices and are excluded from the analysis. The
dashed horizontal line indicates the fitted background level.}
\label{xmmslice}
\end{figure}

In order to determine the extension
of the extended X-ray feature, a slice on the count map containing the pulsar was taken 
(Fig.~\ref{xmmslice}, top) in the direction determined by the azimuthal projection, with a slice
width of $88\arcsec$ and a length of $238\arcsec$ (on slice). A background slice of the same size and
orientation was chosen in a source free region at roughly equal offset to the center of the FOV as
the on slice to ensure equal exposure.  The slices are completely contained within single chips in
the MOS1 and MOS2 cameras and extend $\sim 40\arcsec$ over the edges of neighboring chips in the pn camera.
Profiles of the on slice and background slice are shown in Fig.~\ref{xmmslice} (middle and bottom). 

A point source located just West of the pulsar, \softxmm, is presumably unrelated
to the pulsar due to its soft nature (the hardness ratio R2, comparing the $1-2$\,keV to $0.5-1$\,keV
bands, is $0.12 \pm 0.08$ compared to $0.76 \pm 0.12$ for the source associated to the pulsar).

The slice (Fig.~\ref{xmmslice}, middle) does not exhibit enough statistics to precisely determine
the morphology of the X-ray extension, but the extension appears to consist of a more
compact region near the pulsar position, referred to in this section as the ``compact region'', 
and extending $\sim 40\arcsec$ to the
left, corresponding to the 2XMMi catalog source \psrxmm\ as well as a feature extending from
$\sim -40\arcsec$ to $\sim -150\arcsec$, referred to here as the ``diffuse''
emission region.

A simultaneous fit of the slices was performed, consisting of a fit to the
unrelated soft point source to the West of the pulsar, a Gaussian to the ``compact" region near the
pulsar position, a larger Gaussian to the ``diffuse'' region extending to the East and a constant to
the counts in the background slice. The fit resulted in a diffuse emission centered at $-104\arcsec \pm  18\arcsec$ with a Gaussian width of
$\sigma = 66\arcsec \pm 19\arcsec$ while the compact region was found to be centered at $-10\arcsec \pm 4\arcsec$ with
a width of $16\arcsec \pm  4\arcsec$.
The unrelated point source was fitted with a King profile
\begin{equation}
f(x) = \frac{C}{\left(1+ (\frac{x-x_{0}}{R_{0}})^{2}\right)^{\alpha}},
\end{equation}
with $R_{0} = 4.3\arcsec$ and $\alpha = 1.5$, corresponding to the PSF of the
\textit{XMM-Newton} pn camera at 1.5\,keV and at $\sim 10'$ offset from the center of the
field of view. For the other cameras, the PSF is slightly narrower than this.

The total X-ray extension is found to extend roughly $170\arcsec$ (diffuse center $+ 1\sigma$
width) beyond the pulsar position, however, the tail of the extension may be cut short by the edge of
the pn chip. However, taking an integration region from the edge of the pn chip, to avoid effects of
changing sensitivity across chips, to the pulsar position (avoiding the soft point source to
the west), for a total integration length of $145\arcsec$ gives total on-counts of 950, and total
background counts 689 with the on/off area ratio $\alpha = 1$ for an excess of 261 corresponding to a detection significance of 6.5\,$\sigma$.

\subsection{X-ray spectrum}
\begin{figure}
\centering
\resizebox{\hsize}{!}{\includegraphics{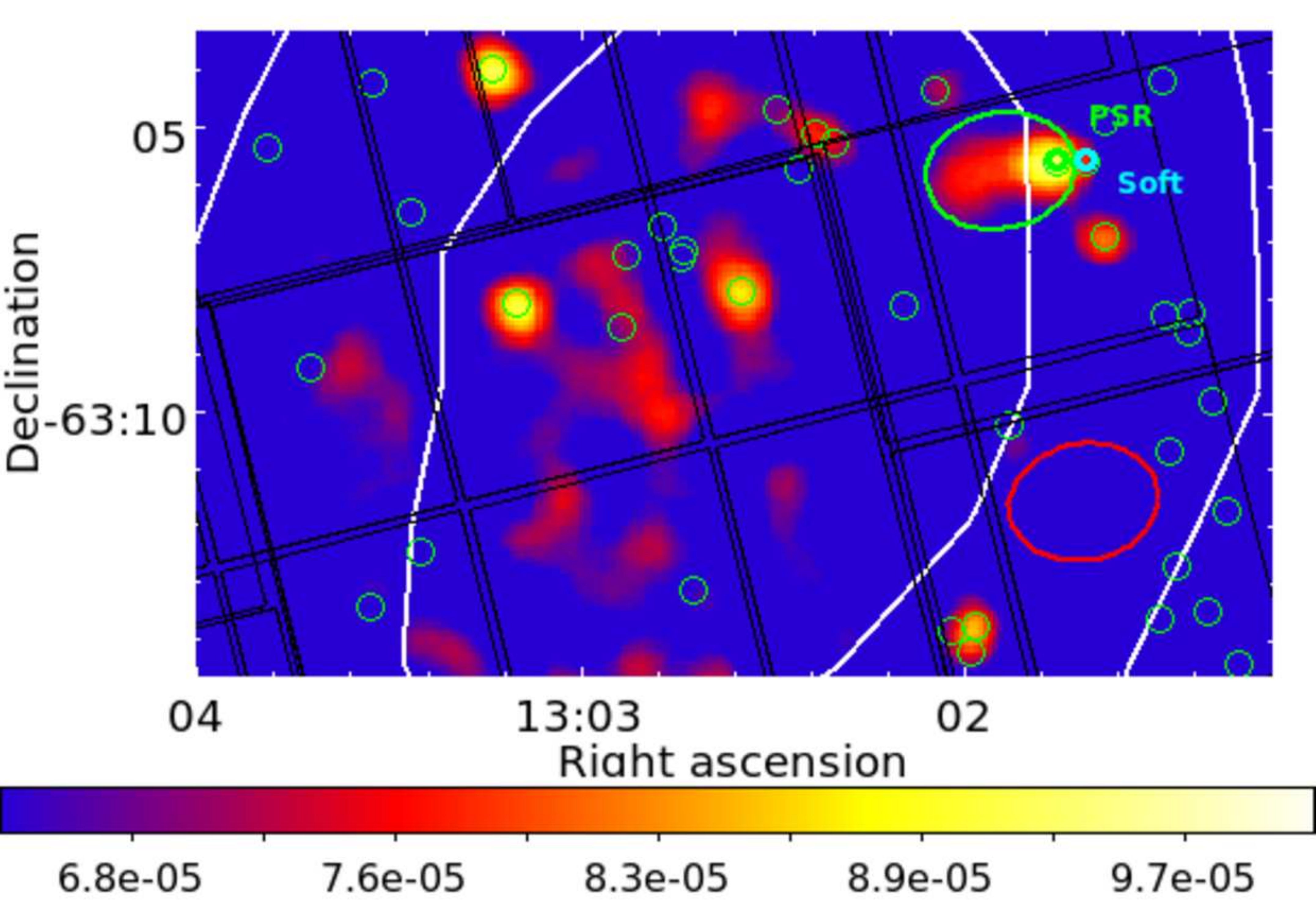}}
\caption{Zoom of Fig.~\ref{xmmslice} to the region around \psrj. 
The region used for extraction of the X-ray spectrum is marked by 
the green ellipse. The red ellipse shows the region used for background determination.
The VHE 14 and 20\,$\sigma$ significance contours are shown in white.}
\label{xmmspec}
\end{figure}

For the spectral extraction, placement of a ring for background determination was not
possible due to multiple nearby sources.  A small elliptical region covering the extension region was taken
and an identical ellipse was used for background extraction (Fig.~\ref{xmmspec}). 
The extraction regions are well contained within single chips for the MOS and pn cameras.  
The spectra were obtained for the three
cameras indepedently and then fit simultaneously. 

The obtained spectrum was fit using the spectral fitting software XSPEC  
with an absorbed power-law model which yielded a column
density $N_{H} =  2.7_{-1.1}^{+1.3}\times 10^{22}\textrm{\,cm}^{-2}$, a photon index
$\Gamma = 2.0_{-0.7}^{+0.6}$, and a flux normalization at 1\,keV of 
$6.2_{-3.8}^{+10}\times 10^{-5}\textrm{\,keV}^{-1}\textrm{\,cm}^{-2}\textrm{\,s}^{-1}$.
The integrated unabsorbed flux in the $2-10$\,keV energy band was found to be
$$
F_{\rm 2-10\,keV} =  1.6^{+0.2}_{-0.4}\times 10^{-13} \textrm{\,erg\,cm}^{-2}\textrm{\,s}^{-1}.
$$

\section{Radio observations}
\begin{figure}
\centering
\resizebox{\hsize}{!}{\includegraphics{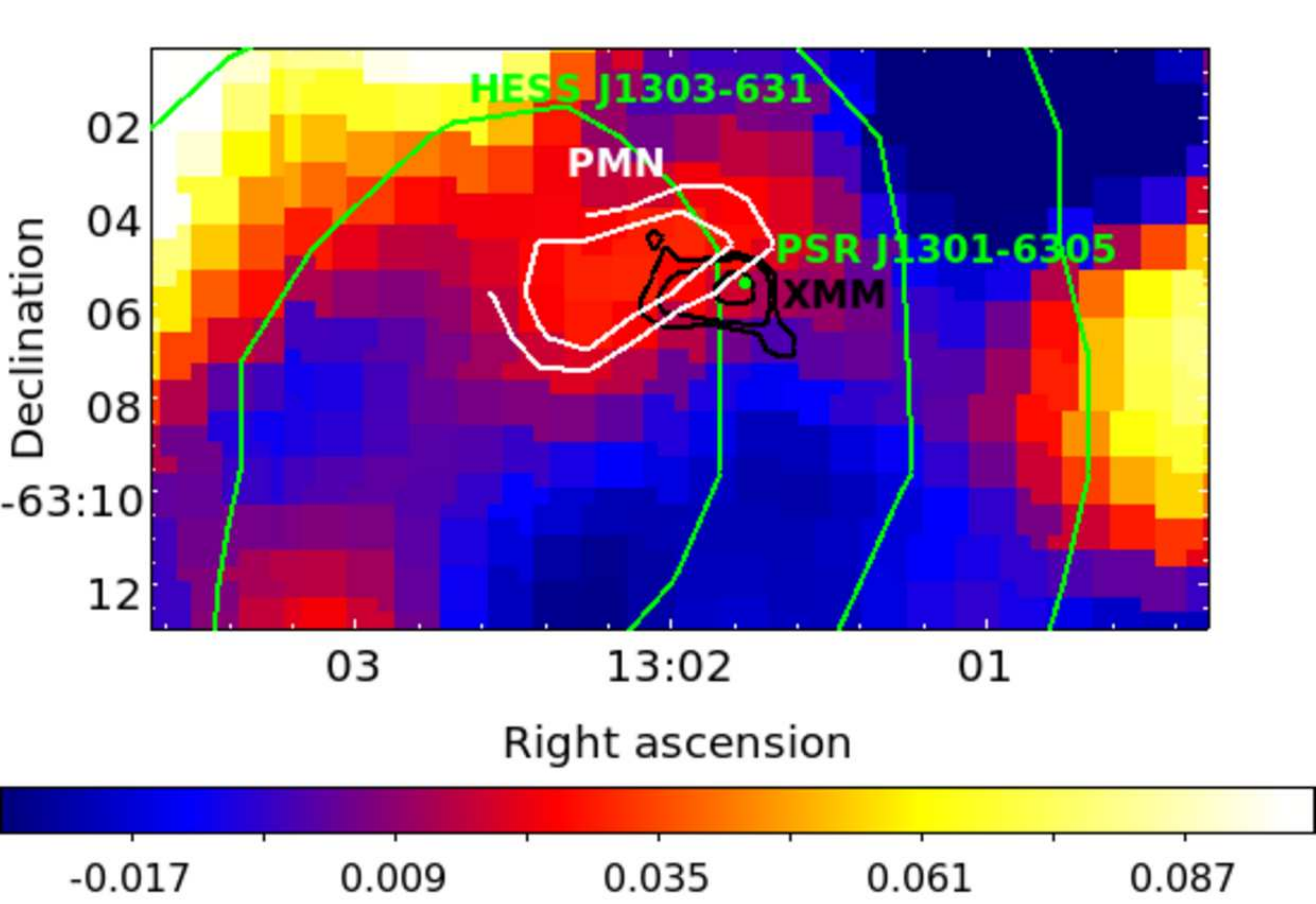}}
\caption{4.85 GHz radio image from the PMN Survey (beam size FWHM $7\arcmin$) in the \hessj\ region.  
The horizontal axis is Right Ascension and the vertical axis is Declination
in J2000.0 coordinates 
and scale is Jy/beam.
H.E.S.S. contours are shown in green, \textit{XMM-Newton} X-ray contours are
shown in black and the radio contours are shown in white.  A radio feature peaks
about 3' East of the pulsar position, just beyond the extended
\textit{XMM-Newton} X-ray source and near the center of the H.E.S.S.
$\gamma$-ray source at a peak value of 0.03 Jy/beam. The apparent position of the radio feature may
be slightly shifted to the North East due to a strong gradient in background from neighbouring sources.}
\label{pmn}
\end{figure}

The region of \hessj\ was covered by a survey of the southern sky by 
the Parkes, MIT and NRAO (PMN) radio telescopes at 4.85 GHz \citep{PMN}.
Calibrated maps were obtained from the NASA SkyView online tool,
shown in Fig.~\ref{pmn}. There is a radio feature
just East of the X-ray nebula and near the peak of the VHE source, the apparent position of which
may be shifted slightly to the North-East due to a strong gradient in the FOV from the strong
unidentified radio sources to the North-East. The feature is found to have a peak flux of $0.03$\,Jy/beam. 
The flux resolution (rms) of the PMN survey is 0.01\,Jy/beam so that the significance of
this feature is only $3\,\sigma$ and is at the detection limit of the survey (and thus not reported
in the catalog). Therefore, the flux is taken as an upper limit. The feature is consistent with the
size of the PSF of the survey ($7\arcmin$ FWHM) in the North-East to South-West direction, but may be
slightly elongated in the North-West to South-East direction, roughly parallel to the
X-ray extension. Since the feature is not significant, no definitive conclusions
about its morphology can be made.

Although it is unclear whether this radio feature does indeed represent
a counterpart of the $\gamma$-ray and X-ray sources, since this lies in
a rather complicated region of the radio sky, the location is promising
due to its similarities with other known PWNe
having a radio peak just beyond the X-ray nebula, as in, for example, 
PSR B1929+10 \citep{psrb1929} and the much smaller scale example of 
the ``Mouse'' PWN \citep{mousethatsoared}. 
Further observations in radio may be able to determine more precisely the morphology
and polarization of the feature in order to establish an association.

\section{Discussion}           
\label{discussion}
\subsection{Multi-wavelength situation}
\begin{figure}
\centering
\resizebox{\hsize}{!}{\includegraphics{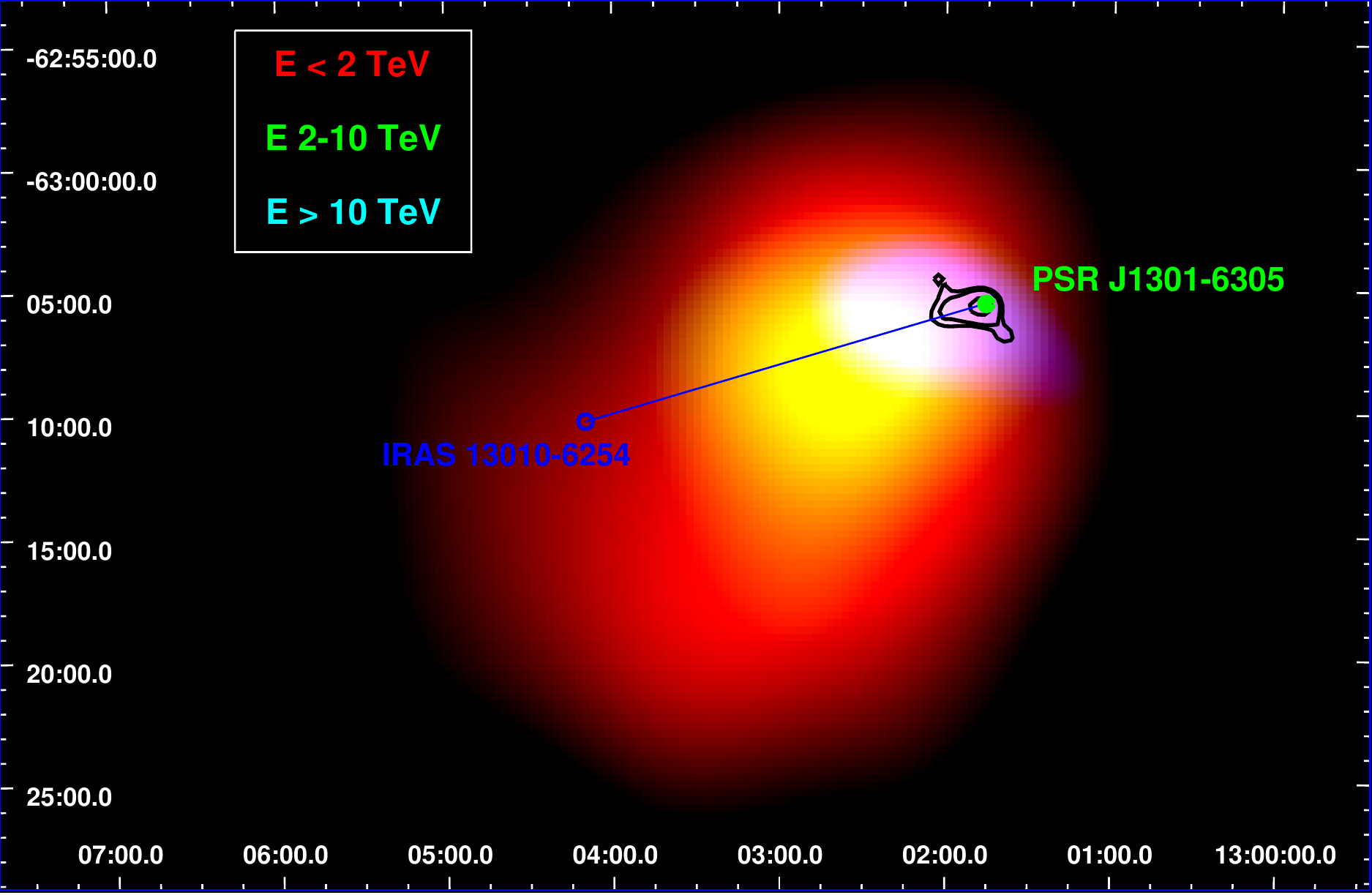}}
\caption{Energy mosaic of HESS J1303-631. 
The horizontal axis is Right Ascension and the vertical axis is Declination
in J2000.0 coordinates. 
Red: E$_{1}$ = (0.84 - 2)\,TeV, Green: E$_{2}$ =(2 - 10)\,TeV and
Blue: E$_{3} >10$\,TeV. The highest energy photons originate nearest the pulsar, \psrj\ (marked by
the green dot).  The visible red corresponds roughly to the 10-$\sigma$ significance contour
of the entire source. 
\textit{XMM-Newton} X-ray contours are shown in black.
A potential birthplace for the pulsar, \iras, as indicated by the X-ray extension, is shown by a
blue circle.}
\label{mosaic}
\end{figure}

Having analysed the morphology and spectra in VHE $\gamma$-rays, X-rays and radio data available for
the region, it is now possible to consider \hessj\ in a full multi-wavelength scenario.
First, an energy mosaic of the VHE emission was created 
using the three smoothed excess images
from Fig.~\ref{edepskymaps}.
These images were overlaid, as shown in Fig.~\ref{mosaic}, 
along with the contours of the extended X-ray PWN.
This energy mosaic is rather reminiscent of the
known \textit{off-set PWN} association HESS\,J1825$-$137 \citep{j1825} where the low-energy 
VHE $\gamma$-ray emission region is quite extended with the pulsar laying
towards the edge of emission and with the higher energy emission more compact and found centered
closer to the pulsar.

\begin{figure}
\centering
\resizebox{\hsize}{!}{\includegraphics{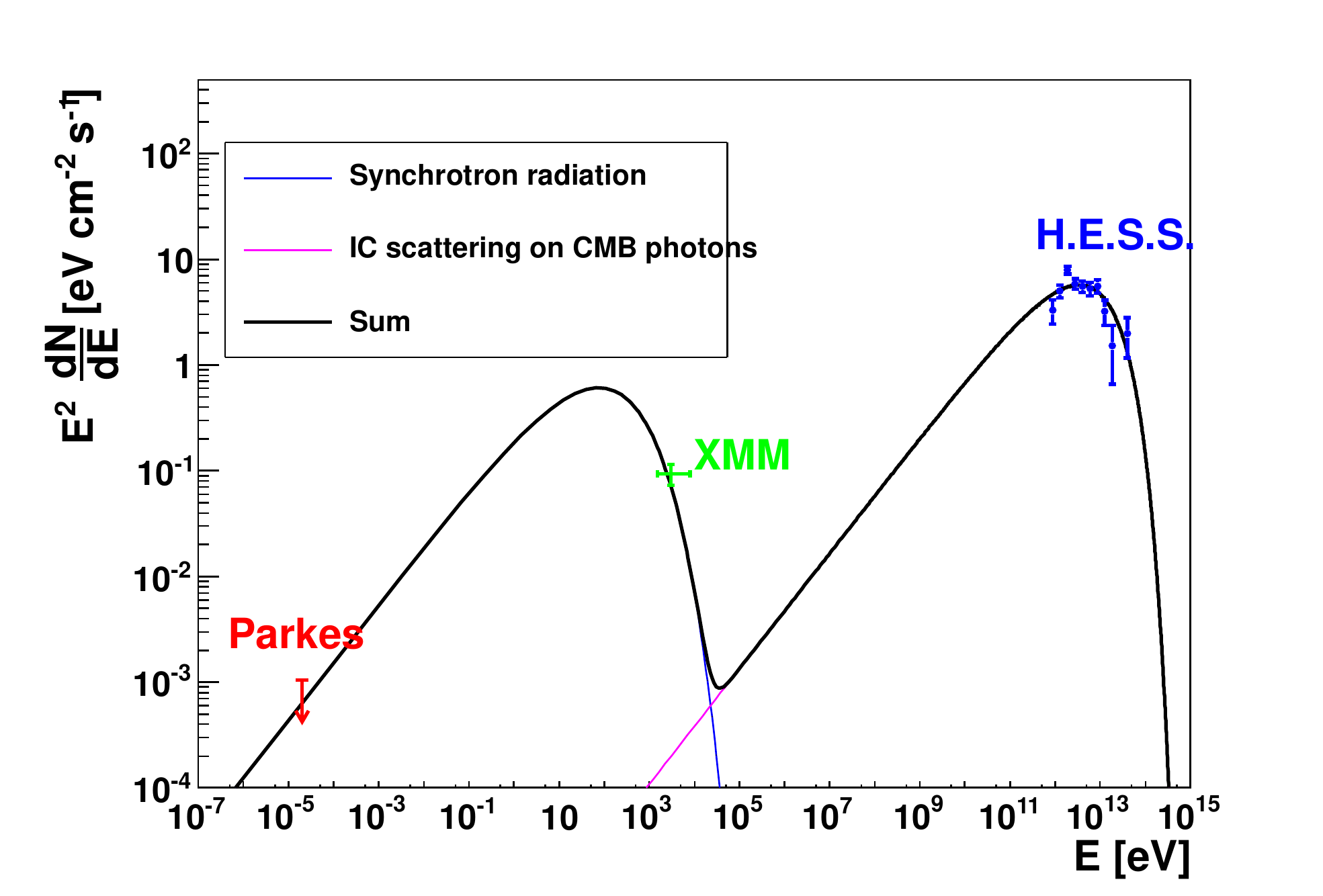}}
\caption{Spectral Energy Distribution of \hessj\ fitted with a simple stationary leptonic model.
The required magnetic field is $\sim 1.4$ $\mu$G.
}
\label{sed}
\end{figure}

Taking the spectra and fluxes obtained in previous sections, it is now
possible to consider the SED of the
source in a PWN scenario. Although a time-dependent model, including the evolution 
of the lepton populations over time, would be required to accurately describe
the emission observed in the various wave bands, for simplicity, and due to the limited number of
multi-wavelength data available, a simple stationary ``one zone'' leptonic model \citep{onezone} was used to fit the
VHE $\gamma$-ray and X-ray spectra as well as the single PMN upper limit in radio (Fig.~\ref{sed}).
The leptonic model assumes that the same electron population, with an energy distribution 
in the form of a single power-law with
an exponential cut-off, creates radio and X-ray emission via synchrotron emission as well as VHE
$\gamma$-rays via inverse Compton (IC) scattering on Cosmic Microwave Background photons. 
Inclusion of IC scattering on
infrared and optical target photons (as obtained from GALPROP \citep{galprop}, assuming 
a pulsar distance of 6.6\,kpc)
had a negligible effect on the model parameters.

The fit of the radio upper limit, and the X-ray and $\gamma$-ray fluxes with
this model yielded an electron spectral index of $\alpha = 1.8^{+0.1}_{-0.1}$, a cut-off
energy of $E_{\rm cut}= 31^{+5}_{-4}$\,TeV, a normalization of 
$K_{e} = 3.7^{+8.1}_{-2.9}\times 10^{6}$\,cm$^{-2}$ 
and an average magnetic field of $1.4^{+0.2}_{-0.2}\,\mu$G,
which is similar to the inferred mean line-of-sight magnetic field strength of $\sim 2\,\mu$G,
as determined from the pulsar's rotation measure \citep{radiopol},
but larger than the magnetic field of $\sim 0.17\,\mu$G predicted by the 
$\gamma$-ray to X-ray luminosity scaling law given in  \cite{onezone}.
The p-value of the fit was 0.02 and the model predicts a total energy in electrons of
$\sim 2.3\times 10^{48}$\,erg. 
It is worth noting
that the resulting model spectrum in the radio band is steeper than
typically observed in PWNe since the X-ray spectral index is not
constrained and the fit of the single electron population is dominated
by the narrow peak in TeV energies. Since
the fluxes at the various energies described by this model are extracted from regions of
differing size, the fitted magnetic field represents only an average and should be 
interpreted with caution.

The differing sizes of the $\gamma$-ray and X-ray emission regions imply the existence of 
differing electron populations so that the entire PWN cannot be accurately modeled by a single
population.  The simple approach presented here, therefore, suffers from the caveat
that a model with two electron populations could reproduce the observed spectra with a significantly
different magnetic field than obtained with a one zone model.
Incorporating a strong cutoff in
an older electron population at high energies would suppress the X-ray synchrotron emission, even in
the face of a much higher magnetic field, and still reproduce the VHE peak. 
Indeed, the higher energy synchrotron emitting electrons may
have been effectively extinguished precisely because of the high magnetic field. The morphology in
VHE $\gamma$-rays shows no evidence of a distinct break in the populations of electrons
caused by passage of an SNR shock, but rather appears to show a more continuous transition from
lower to higher energies in VHE $\gamma$-rays and on up to the highest energy synchrotron X-ray emitting electrons
closer to the pulsar. This would imply a continuous transition from older to younger electrons
which may require not a two zone electron model, but a continuously changing population making modeling
quite difficult. Due to scant spectra available at lower energies, the precise details of the
electron populations cannot be distinguished, and this first order approximation model serves as
a starting point for future studies and searches.

The VHE $\gamma$-ray morphology presented here favors the 
association of \hessj\ with the high spin-down power pulsar 
\psrj, which, on energetic grounds, is the
only known association which can explain the TeV emission.  Additionally, the detection of an
extended asymmetric X-ray nebula in combination with a hint of a radio counterpart of the 
PWN seen in PMN observations, strengthens this association further.

\subsection{Distance to the source}
As stated before, the distance of 6.6\,kpc to \psrj\ is based on the
dispersion measure using the model of electron distribution in the Galaxy, a method which is often considered unreliable.
For example, the ``Mouse'' pulsar has been
argued to be at a distance roughly twice that determined by its dispersion measure based on the
ratio of neutral hydrogen atoms to free electrons along the line of sight
of $N_{H}/$DM = 85
which is much higher than the values seen for all other X-ray detected pulsars, for which typically
we observe $N_{H}/$DM $\approx 5 - 10$ \citep{mousethatsoared}.
For \psrj, using the column density obtained here, we have $N_{H}/$DM $\sim 23$, one of the highest
known $N_{H}/$DM ratios among PWNe, which could imply that the distance obtained from DM is an underestimate for this source as well.

On the other hand, the star formation region \iras, if considered as a potential birthplace of the
pulsar, provides an alternative estimation of the distance. The direction of the X-ray extension is
found to be within 1\,$\sigma$ from the direction of the star formation region, \iras, located at a position angle of $106.3^{\circ}$, the only other identified object within the VHE emission
region besides lower energy pulsars and stars. 
\iras\ has been identified as a point source in GeV $\gamma$-rays in Fermi observations by
\cite{irasgamma} which could indicate the presence of evolved massive stars, which are
the progenitors to pulsars,
with colliding winds, or the presence of an
SNR within the star forming region, or potentially a counterpart to a part of the VHE source
given the bulge in the VHE significance contours at the position of the star forming region
(see Fig.~\ref{annulus}, top). As the only star formation region within more than a degree of the pulsar,
and in the absence of another plausible SNR association, this provides a plausible candidate
for the birthplace of the pulsar.  
This could be similar to the case of the X-ray feature G359.95$-$0.04 which was identified by
\cite{pwntostarforming} as a PWN and found extending in the direction of the young stellar complex IRS 13, which was
suggested as a possible birthplace for the yet undetected pulsar. 

\iras\ has a kinematic velocity of $V_{\rm SLR} = (33.4 \pm 3.2)$ km/s 
\citep{iraspoints}, which corresponds to a distance of $\sim 12.4-12.9$\,kpc
using the circular Galactic rotation model of \cite{galrotation}
updated with the Galactic structure parameters of \cite{galstruct}. This is nearly double the
distance of 6.6\,kpc based on DM, placing the source close to the edge of the Galaxy. This kinematic
distance is corroborated by the measure of the column density from X-rays, which is 
larger than the total integrated Galactic HI column density in that direction
of $1.9\times 10^{22}$\,cm$^{-2}$ \citep{hi}.

If the pulsar was born in \iras\ then it would
have traveled $0.28^{\circ}$ or $\sim 62$ pc, implying a very high transverse velocity of $\sim
5,000$ km/s if the characteristic age of 11 kyr is taken as the true age.  This age estimate is,
however, often considered to be unreliable and the true ages may differ by a factor of 2-3.
The true age of the pulsar, assuming constant braking index, is given as \citep{pulsarsbook}
\begin{equation}
\tau = \frac{P}{(n-1)\dot{P}}\left[1- \left(\frac{P_{0}}{P}\right)^{n-1}\right]
\end{equation}
The characteristic age $\tau_{c}$ is calculated assuming a braking index of $n=3$ (i.e. pure electro-magnetic braking)
and that the birth period, $P_{0}$ is much less than the current period, $P$.  
The braking index has only been reliably measured for a handful of young pulsars \citep{brakindex}
and was found to be less than 3 in every case, with the extreme case of  with $n= 1.4 \pm 0.2$ \citep{velan}
implying an age 5 times greater than predicted by $\tau_{c}$ if the assumption of $P_{0}<<P$ still holds,
or PSR J1734$-$3333 with an index of $n= 0.9 \pm 0.2$ \citep{j1734} implying an age which cannot be calculated
with the above formula.
\cite{qvf} calculated the effects of quantum vacuum friction on the spindown of pulsars and found
a braking index decreasing as $1 - (1-n_{0})e^{-At}$ with 
$A=\ddot{P}/\dot{P} + \dot{P}/P$ for period $P$ and predict 
the braking index of the Crab pulsar at birth of just below 3 and that it will fall
to $\sim 2$ in the next 2\,kyr from its current value of 2.5.
\cite{oldpsrs} proposed that  ``characteristic ages greatly underestimate the true ages of pulsars''
based on proper motion measurements of PSR~B1757-24.
On the other hand,  \cite{CTB80} used proper motion measurements to show that the pulsar PSR~B1951+32 
is likely $\sim 40\%$ younger than its characteristic age, implying a non-negligible birth period.

The very large ``darkness ratio'' of
$\gamma$-ray to X-ray luminosity (in the $1-30$\,TeV and $2-10$\,keV bands respectively)
for this source of 156, 
makes this the darkest identified PWN to date 
(the darker HESS J1702-420, darkness ratio 1,500, is now believed to be an SNR, \cite{SNRJ1702}).
This could imply a relatively old age for \psrj.
\cite{gammax} considered the darkness ratios 
of PWNe and PWNe candidates detected by H.E.S.S., and found a 
logarithmic scaling with the characteristic age of the
assumed associated pulsar.  
This scaling law predicts a darkness ratio of 6 for a pulsar with characteristic age 11\,kyr,
whereas the darkness ratio measured here would predict a
characteristic age of 48\,kyr, more than 4 times older.

A factor of 3 to 5 in the age of \psrj\  would bring the pulsar velocity down to $\sim 1,600$ to $1,000$\,km/s.
This high velocity is not unreasonable given the two component pulsar velocity model by
\cite{psrvelo2002} which predicts $\sim 15\%$ of all pulsars to have a space velocity greater than $1,000$ km/s,
but would place \psrj\ among the fastest known pulsars, including the Guitar Nebula
pulsar (PSR~B2224+65), PSR~B1953+50, PSR~B1800-21, PSR~B1757-24 and 
PSR~B1610-50 all believed to have a velocity of $\gtrsim$\,1,600\,km/s (see
\cite{psrvelo1998,snrsNyoungpsrs}). 
In the case that the pulsar is much older than 11 kyr, IC
cooling may play an important role for the oldest electrons, i.e. those created nearest the place of birth,
leading to strong energy-dependent morphology as observed here.

Adopting the kinematic distance of \iras, the integrated $\gamma$-ray luminosity would
represent about 28\% of the current spin-down luminosity of the pulsar 
($\dot{E}_{12.6} = \dot{E}/4\pi(12.6\textrm{ kpc})^{2} = 8.9\times 10^{-11}$\,erg\,cm$^{-2}$\,s$^{-1}$).
This $\gamma$-ray conversion efficiency is higher than for typical PWNe ($\lesssim10\%$ for PWNe
with known $\gamma$-ray and X-ray luminosities, \cite{gammax}). 
However, high $\gamma$-ray conversion efficiency
may not be unreasonable considering the very high ``darkness ratio'' of 
for this source of 156,
which implies low synchrotron losses.
Thus, an association of \psrj\ to the star forming region \iras\ cannot be ruled out,
while the association is supported by the direction of the X-ray trail, the high
ratio of $N_{H}/$DM, the large absolute value of $N_{H}$ consistent with the entire integrated galactic column density in that direction,
the possible bump seen in the TeV emission at the location of \iras, the recent detection of \iras\ by Fermi 
as well as the high ratio of $\gamma$-ray to X-ray luminosity.
This larger distance may also help explain the difficulty 
of detecting counterparts at other wavelengths.

\section{Conclusions}           
PWNe now appear to constitute the largest class of Galactic VHE $\gamma$-ray
emitters. The first dark source, and considered ``prototypical'' dark source, 
TeV J2032+4130 discovered by the HEGRA collaboration \citep{TeV2032}, 
was recently found to be ``not-so-dark'',
after deeper X-ray and radio observations have revealed 
weak but significant counterparts \citep{notsodark} and the identification
of a pulsar first in $\gamma$-rays \citep{fermibrightsource}
and then in radio \citep{radioJ2032}.
The work presented here has successfully identified energy-dependent morphology in VHE $\gamma$-rays
as well as an X-ray PWN counterpart of
\hessj, which now appears also to belong to the ``not-so-dark'', or ``synchrotron under-luminous''
class of VHE $\gamma$-ray sources having peak synchrotron energy fluxes that are much lower
than the peak fluxes in the VHE regime.
The observations presented here support the
interpretation of this source as a large cloud of electrons, accelerated by the pulsar,
which emit $\gamma$-ray radiation through the IC mechanism.  
These electrons can have an IC emission
lifetime of the order of the pulsar age, and can, therefore, reflect the total energy
output of the pulsar since birth, while the X-ray part of the PWN, generated by higher
energy synchrotron emitting electrons with a much shorter interaction time, 
decreases rapidly in time
and reflects only the more recent spin-down power of the pulsar \citep{darkpwn}.
While an association
of the pulsar with the star formation region \iras\ is far from clear, 
it has been shown that it at least cannot be ruled out. The larger distance obtained
from \iras\ may explain the very high absorption column density obtained from X-rays.
The pulsar's true age, distance and origin remain open questions, as well as the
details of the underlying electron populations that are responsible for the multi-wavelength
emission.  Current efforts to extend the radio and X-ray measurements of this source
will be crucial for a deeper understanding of the processes at play.

Many other extended Galactic $\gamma$-ray sources which were previously unidentified
are also finding associations with pulsars and PWNe as this class of sources continues to expand.
The results obtained here also support the hypothesis that this ``not-so-dark''
source may be understood in the context of very low magnetic field, possibly in combination with a large 
distance to the source, causing relative extinction of the X-ray
counterpart.

\begin{acknowledgements}
The support of the Namibian authorities and of the University of
Namibia in facilitating the construction and operation of H.E.S.S.\ is
gratefully acknowledged, as is the support by the German Ministry for
Education and Research (BMBF), the Max Planck Society, the French
Ministry for Research, the CNRS-IN2P3 and the Astroparticle
Interdisciplinary Programme of the CNRS, the U.K. Particle Physics and
Astronomy Research Council (PPARC), the IPNP of the Charles
University, the South African Department of Science and Technology and
National Research Foundation, and by the University of Namibia. We
appreciate the excellent work of the technical support staff in
Berlin, Durham, Hamburg, Heidelberg, Palaiseau, Paris, Saclay, and in
Namibia in the construction and operation of the equipment.
M. Dalton acknowledges the support of the European Research Council (ERC-StG-259391).
\end{acknowledgements}

\bibliographystyle{aa}


\end{document}